\documentclass[journal]{IEEEtran}
\IEEEoverridecommandlockouts
\usepackage{cite}
\usepackage{amsmath,amssymb,amsfonts}
\usepackage{algorithmic}
\usepackage{graphicx}
\usepackage{textcomp}
\usepackage{xcolor}
\usepackage{glossaries}
\usepackage{bm}
\usepackage{caption}
\usepackage{subcaption}
\usepackage{hyperref}
\usepackage{verbatim}

\def\BibTeX{{\rm B\kern-.05em{\sc i\kern-.025em b}\kern-.08em
    T\kern-.1667em\lower.7ex\hbox{E}\kern-.125emX}}
    
\newacronym{NLL}{NLL}{negative-log-likelihood}
\newacronym{NF}{NF}{normalizing flow}
\newacronym{NLM}{NLM}{nonlocal means}
\newacronym{OBNLM}{OBNLM}{Optimized Bayesian nonlocal means}
\newacronym{PSNR}{PSNR}{peak signal-to-noise ratio}
\newacronym{SSIM}{SSIM}{Structural similarity index measure}
\newacronym{PICMUS}{PICMUS}{Plane-wave  Imaging Challenge in Medical UltraSound}
\newacronym{CUBDL}{CUBDL}{Challenge on ultrasound beamforming with deep learning}
\newacronym{PDF}{PDF}{probability density function}
\newacronym{MAP}{MAP}{maximum-a-posteriori}

\newcommand{\x}[0]{\bm{x}}
\newcommand{\y}[0]{\bm{y}}
\newcommand{\z}[0]{\bm{z}}
\newcommand{\n}[0]{\bm{n}}

\DeclareMathOperator*{\argmin}{arg\,min\;}

\begin{document}

\title{Ultrasound Speckle Suppression and Denoising using MRI-derived Normalizing Flow Priors}

\author{
Vincent~van de Schaft
and Ruud J.G. van Sloun, ~\IEEEmembership{Member,~IEEE}.

\thanks{
V. van de Schaft and R.J.G. van Sloun are with the department of Electrical Engineering, Eindhoven University of Technology, Eindhoven, The Netherlands. (e-mails: v.v.d.schaft@student.tue.nl, r.j.g.v.sloun@tue.nl). R.J.G. van Sloun is also with Philips Research, Eindhoven.}
}

\maketitle

\begin{abstract}
Ultrasonography offers an inexpensive, widely-accessible and compact medical imaging solution. However, compared to other imaging modalities such as CT and MRI, ultrasound images notoriously suffer from strong speckle noise, which originates from the random interference of sub-wavelength scattering. This deteriorates ultrasound image quality and makes interpretation challenging. We here propose a new unsupervised ultrasound speckle reduction and image denoising method based on maximum-a-posteriori estimation with deep generative priors that are learned from high-quality MRI images. To model the generative tissue reflectivity prior, we exploit normalizing flows, which in recent years have shown to be very powerful in modeling signal priors across a variety of applications. To facilitate generaliation, we factorize the prior and train our flow model on patches from the NYU fastMRI (fully-sampled) dataset. This prior is then used for inference in an iterative denoising scheme. We first validate the utility of our learned priors on noisy MRI data (no prior domain shift), and then turn to evaluating performance on both simulated and in-vivo ultrasound images from the \gls{PICMUS} and \gls{CUBDL} datasets. The results show that the method outperforms other (unsupervised) ultrasound denoising methods (\gls{NLM} and \gls{OBNLM}) both quantitatively and qualitatively. 
\end{abstract}


\section{Introduction}
\noindent Ultrasound (US) imaging offers a low-cost and widely accessible diagnostic imaging solution, with form factors becoming ever more compact and handheld. This portability and cost-effectiveness enables point-of-care imaging at the bedside, in emergency settings, rural clinics, and developing countries. US sees increasing adoption across many medical specialties, ranging from cardiology to obstetrics and oncology. Unfortunately, US image quality suffers from many sources of image degradation, one of which being speckle noise. The resulting pseudo-random fluctuations in intensity obscure important image features and make US images hard to interpret. 


Many previous works have found ways to combat speckle noise. The Lee filter \cite{lee_filter} uses local statistics to denoise individual pixels in the presence of multiplicative and/or additive noise. Later methods include the Fost-filter, the Kuan-filter, and speckle reducing anisotropic diffusion \cite{frost_filter, kuan_filter, srad}. The nonlocal-means algorithm uses statistics from patches that are similar to the pixel to realize more smoothing while retaining sharp edges \cite{nlm}. This approach was later adapted specifically for combating ultrasound speckle with the \gls{OBNLM}-algorithm \cite{obnlm,coupe2008bayesian}. 

More recently, inspired by the success of deep learning in computer vision, deep neural network based approaches have been proposed for US denoising. These data-driven methods act as regressors that learn a parameterized mapping from images with speckle to corresponding clean images \cite{structural_convolutional, 5_NN}, or from raw channel data (i.e. before beamforming) to clean images \cite{NN_beamforming}. Deep networks for US image quality improvement are typically based on \textit{discriminative} convolutional architectures (such as the well-known U-Net), and often resort to pairs of clean and artificially corrupted images obtained via simulation for training \cite{NN_beamforming,vedula2017towards,feng2020ultrasound}. Unfortunately, including real US data for improved fitting of the measurement data distribution is challenging due to lack of access to clean target US images. Some works therefore aim at acceleration of existing yet computationally demanding methods that provide such targets \cite{luijten2020adaptive,dietrichson2018ultrasound}, or resort to domain adaptation approaches \cite{tierney2020domain}. A second disadvantage of discriminative methods (especially when trained on simulations) is their sensitivity to out-of-distribution measurement statistics, such as changes in the noise statistics or system point-spread-function characteristics. The alternative, using deep \textit{generative} models as image priors for US image quality improvement has to the best of our knowledge thus far not been explored.

In this paper, we propose to use deep generative models based on Normalizing Flows (NFs) in combination with Maximum-a-Posteriori (MAP) estimation for US image denoising and speckle suppression. NFs \cite{realnvp, nice, glow, flow++, overview1} are generative models which have been shown to be very effective in modeling complex image priors for sample generation, solving inverse tasks such as image inpainting and denoising \cite{asim2020invertible,wei2021deep}, as well as in density estimation \cite{maf}. We here use its ability to explicitly model density functions to fit a high-quality target image distribution, which in turn enables performing US image despeckling/denoising via MAP optimization under a Gaussian likelihood model. To train a high-image-quality anatomical prior, we fit our generative model on patches of (by nature) speckle-free MRI data, sampled from the NYU fastMRI dataset\cite{fastmri1, fastmri2}. Our method thus does not rely on pairs of noisy and speckle-free US images to train a direct estimator, but instead performs unsupervised reconstruction via an explicit generative prior trained a-priori.  

We show that our MRI-derived image priors are very effective at denoising US images across a series of experiments, ranging from simulations with artificially-generated speckle noise to \textit{in-vivo} US images, thereby outperforming strong traditional US speckle reduction algorithms. \\

Our main contributions are as follows:
\begin{itemize}
    \item We formulate the US de-speckling problem as a MAP optimization problem under a deep generative image prior. 
    \item We propose to learn this image prior from high-quality MRI data using NFs, and factorize the distribution to model local tissue structure (patches) rather than high-level anatomical semantics. 
    \item We propose a method for automatic sampling of structurally informative patches from full MRI volumes to learn improved priors for downstream MAP estimation. 
\end{itemize}

The remainder of this paper is organized as follows. Section~\ref{section:background} provides the reader with some background on ultrasound speckle and normalizing flows for deep generative modeling. In sections~\ref{section:methods} and \ref{section:experiments} the methodology, algorithms and the performed experiments are explained, respectively. The results are presented in section~\ref{section:results}. Finally, in sections~\ref{section:discussion} and \ref{section:conclusion}, the results are discussed and conclusions derived. 

\section{Background}
\noindent In this section we provide the reader with some background on the ultrasound acquisition process, leading up to the origin of speckle noise and its characteristics (\ref{section:background_ultrasound}). We then turn to describing \gls{NF}s and their use for flexible generative density modelling (\ref{section:background_NF}).
\label{section:background}
\subsection{Ultrasound speckle}\label{section:background_ultrasound}
\noindent Brightness-mode (B-mode) US aims at imaging the spatial distribution of backscatter intensities in tissue. This is achieved by first transmitting a short ultrasonic pulse (typically in the range of 1-10MHz), probing the tissue, and then beamsteering an array of reiver elements to each individual pixel (pixel-based dynamic receive beamforming) to estimate the local tissue reflectivity (or backscatter intensity). The spatial selectivity of the beamsteering operation (i.e. spatial filtering), impacts the resolution (main-lobe width and pulse length) and contrast (side- and grating-lobes) of the system. 
The nature of this pulse-echo experiment causes so called speckle noise in the resulting image: the result of constructive and destructive interference between reflections originating from many sub-wavelength spaced scatters at pseudo-random locations. This summation of phasors with random phase and amplitude within one resolution cell yields a strong multiplicative noise component in the magnitude domain, that is spatially correlated by the system resolution. A simple convolutional model of speckle noise is thus \cite{yu2002speckle}: 
\begin{equation}
    \y(r_x,r_y) = \left\lbrace \x(r_x,r_y)\cdot \n(r_x,r_y)\right\rbrace*h(r_x,r_y),
\end{equation}
where $\y$ is the speckle-corrupted image, $\x$ is the clean backscatter magnitude image, $\n$ is an i.i.d. multiplicative (Rayleigh) noise component, $*$ denotes a spatial convolution, and $h$ is the radiofrequency(RF)-modulated point spread function. Speckle-corrupted RF US images are then envelope-detected and log-compressed before display. Empirical models for log-compressed, envelope-detected US data often assume speckle noise models of the form $u(r_x,r_y) = v(r_x,r_y) + \sqrt{v(r_x,r_y)}\eta(r_x,r_y)$, with $u(r_x,r_y)$ the noise-corrupted image, $v(r_x,r_y)$ the clean image, and $\eta(r_x,r_y)$ zero-mean Gaussian noise with some standard deviation \cite{coupe2008bayesian}. Importantly, speckle noise is thus signal dependent. 

\subsection{Deep generative models and normalizing flows}
\label{section:background_NF}
\noindent Deep generative modeling aims to learn the structure of data by fitting complex parametric distributions to training examples. Such distributions should satisfy $\int_{\x} p_X(\x) = 1$. For arbitrary neural networks this does not hold and enforcing this quickly becomes intractable for high dimensional problems. This problem is addressed by either optimizing an evidence lower bound of the likelihood under arbitrary neural architectures (VAEs), implicitly optimizing likelihood through adversarial training (GANs), or more recently through the use of architectures based on invertible bijective mappings: normalizing flows. Normalizing flows are a class of models that learn to transform a complex high-dimensional data distribution into a simple base distribution through a series (flow) of such invertible transformations. This base distribution is often chosen to be a Gaussian distribution (hence the name \textit{normalizing flow}), enabling straightforward sampling. Deep generative models based on normalizing flows have the advantage of enabling directly likelihood estimation of data samples, 
\\
To achieve this, normalizing flows leverage the change of variables formula. Let $X\in\mathbb{R}^N$ be a random variable from which we wish to model the probability density function $p_X(x)$. Let $f_\theta(x)$ be a differentiable, invertible transformation parameterized by $\theta$. 
$$Z = f_\theta(X)$$
where $X\sim p_X(\x)$ and $Z\sim p_Z(\z)$. The change of variables formula can be used to determine the probability density function of random variable $X$ as a function of the probability density function of $Z$ as
\begin{equation}
    p_X(\x) = p_Z(f_\theta(\x))\cdot |\det J_{\x}(f_\theta(\x)) |
    \label{eq:change_of_variables}
\end{equation}
where $J_x(f_\theta(x))$ is the Jacobian of the transformation. The first term, $p_Z(f_\theta(\x))$, is the probability density function of the base distribution evaluated at $\z$. The second term, $|\det J_{\x}(f_\theta(\x))|$, corrects for the change in volume caused by the transformation.
The corresponding \gls{NLL} is given by
\begin{equation}
-\log p_X(\x) = -\log p_Z(\z) - \log|\det J_{\x}(f_\theta(\x))|,
\end{equation}
which can be directly optimized on a training dataset. 
Once the transformation has been learned it provides a model of $p_X(\x)$. To evaluate the likelihood of a sample $p_X(\x)$ one can compute $z = f_\theta(\x)$ and evaluate the probability density function of the base distribution at $\z$, combine it with the determinant of the Jacobian using (\ref{eq:change_of_variables}).
Sampling from $p_X(\x)$ is also straightforward. One can draw a sample $\z$ from the simple base distribution and transform it to the data distribution using the inverse transformation
\begin{equation}
\x = f_\theta^{-1}(\z). 
\end{equation}

\begin{figure*}[t!]
    \centering
    \includegraphics[trim={0cm 20.5cm 4cm 0},clip,width=0.75\linewidth]{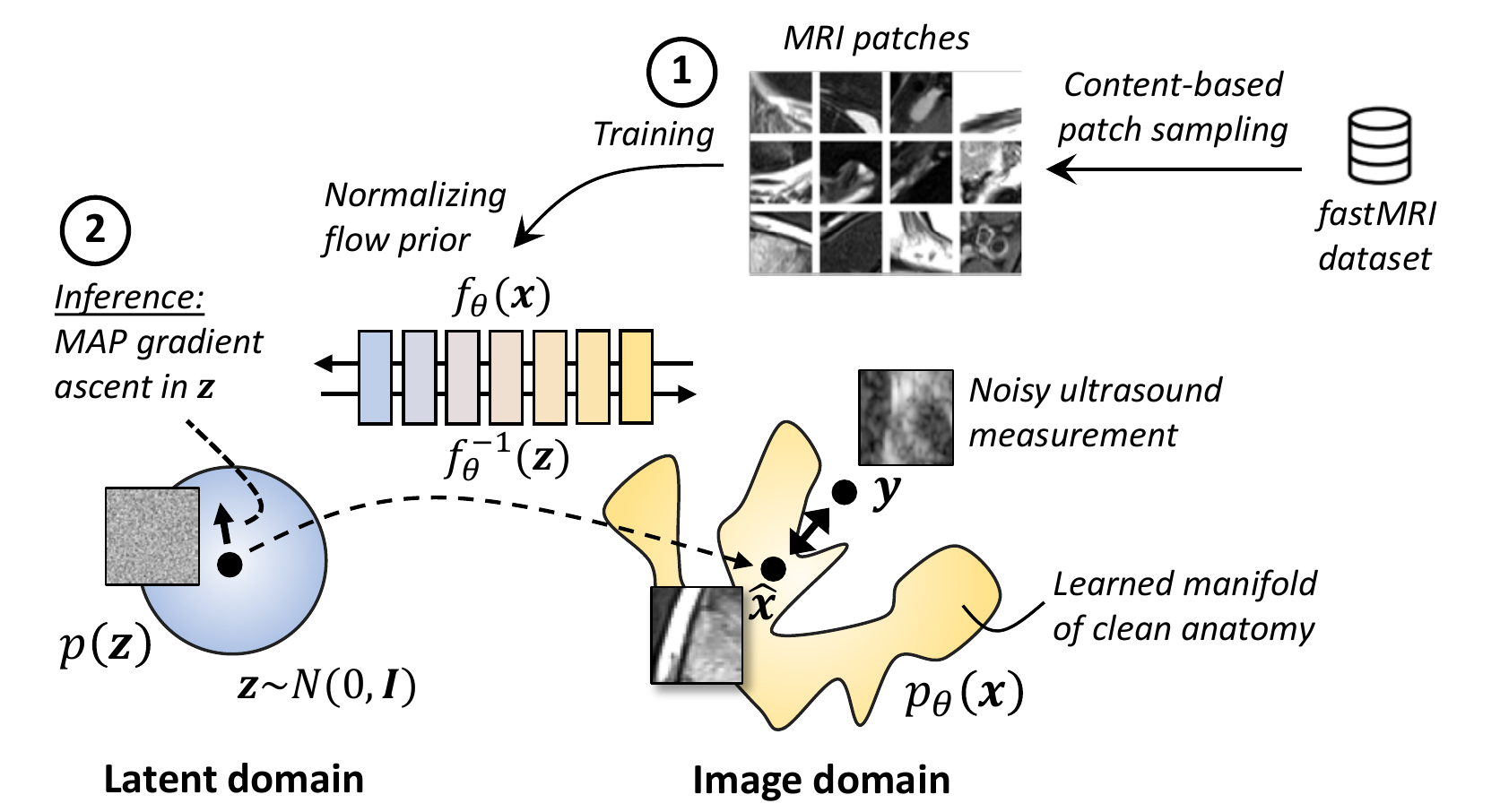}
    \caption{An overview of the proposed approach. We first (1) train a normalizing flow prior on patches sampled from (by nature) speckle-free anatomical MRI images. We then (2) use this prior for MAP inference on noisy ultrasound measurements, by performing gradient ascent on the posterior in latent $\z$ space, minimizing the $\ell_2$ distance between noisy measurements and images lying on the learned manifold of clean anatomical images.} 
    \label{fig:overview}
\end{figure*}

\section{Methods}
\label{section:methods}
\noindent In this section the proposed method and experimental details are provided. The denoising algorithm is explained in section \ref{section:methods_overview}. Neural network architecture of the prior model as well as the training data are discussed in section \ref{section:methods_prior}. 

\subsection{MAP denoising using generative priors}
\label{section:methods_overview}
\noindent We formulate the task of recovering the clean image $\x$ from a measured, noisy, image $\y$ as a MAP optimization problem:
\begin{align*}
   \hat{\x} &= \arg\max_{\x} p_{X | Y}(\x\;|\;\y) \\ &= \arg\max_{\x}  p_{Y|X}(\y\;|\;\x)\cdot p_X(\x).
\end{align*}
Under a given deep generative \gls{NF} prior with normalized hidden space $\z$, we then perform optimization in the log-$z$-domain, such that:
\begin{align*}
    \hat{\z} &= \arg\max_{\z} \log p_{Z|Y}(\z\;|\;\y) \\ &= \arg\max_{\z} \left\lbrace \log p_{Y|X}(\y\;|\;f_\theta^{-1}(\z))+\log p_Z(\z) \right\rbrace ,
    \label{eqn:MAP1}
\end{align*}
where  $f^{-1}_\theta(\z)$ is the inverse transformation of $f_\theta(\z)$, i.e. the generative direction. We here approximate the measurement process as a simple Gaussian likelihood model for the log-compressed envelope detected US images, such that:
\begin{equation}
    \y = f^{-1}_\theta(\z)+ n,
\end{equation}
where $n$ follows an i.i.d Gaussian distribution. Since $p_Z(\z)$ by design also follows a normal distribution, we can rewrite the above optimization problem as:
\begin{equation}
    \hat{\z} = \argmin_{\z} || f^{-1}_\theta(\z) - \y||_2^2 + \lambda||\z||_2^2,
    \label{eqn:MAPfinal}
\end{equation}
where $\lambda$ is a parameter that depends on the assumed noise variance of $n$. Intuitively this means that the algorithm places more trust in the prior when noise variance is high. 

In this work we perform optimization using gradient descent with the Adam optimizer, making use of the Pytorch Autograd functionality to automatically compute gradients. The gradient step size is decayed each time the loss reaches a plateau. We finally recover the clean image as $\hat{\x}=f^{-1}_\theta(\hat{\z})$. Fig.~\ref{fig:overview} gives an overview of the proposed approach. 

\subsection{Prior factorization and MRI content-based patch sampling}
\noindent The design of the prior distribution clearly $p_X(\x)$ has a strong impact on the effectiveness of our strategy, both in terms of noise/speckle suppression as well as generalization to unseen anatomical instances. To achieve the latter we factorize our prior distribution to only consider the joint distribution of pixels within a local image patch of $14\times14$~mm ($32\times 32$ pixels), retaining independence across larger spatial distances and thereby per design not modeling high-level semantics that characterize particular antomical (or even pathological) images. Note that for final image-level denoising according to \eqref{eqn:MAPfinal}, we perform parallel inference (250 patches in parallel) via gradient checkpointing \cite{gradient_checkpointing} and then stitch the patches together. 

\begin{figure}
    \centering
    \includegraphics[width=.9\linewidth]{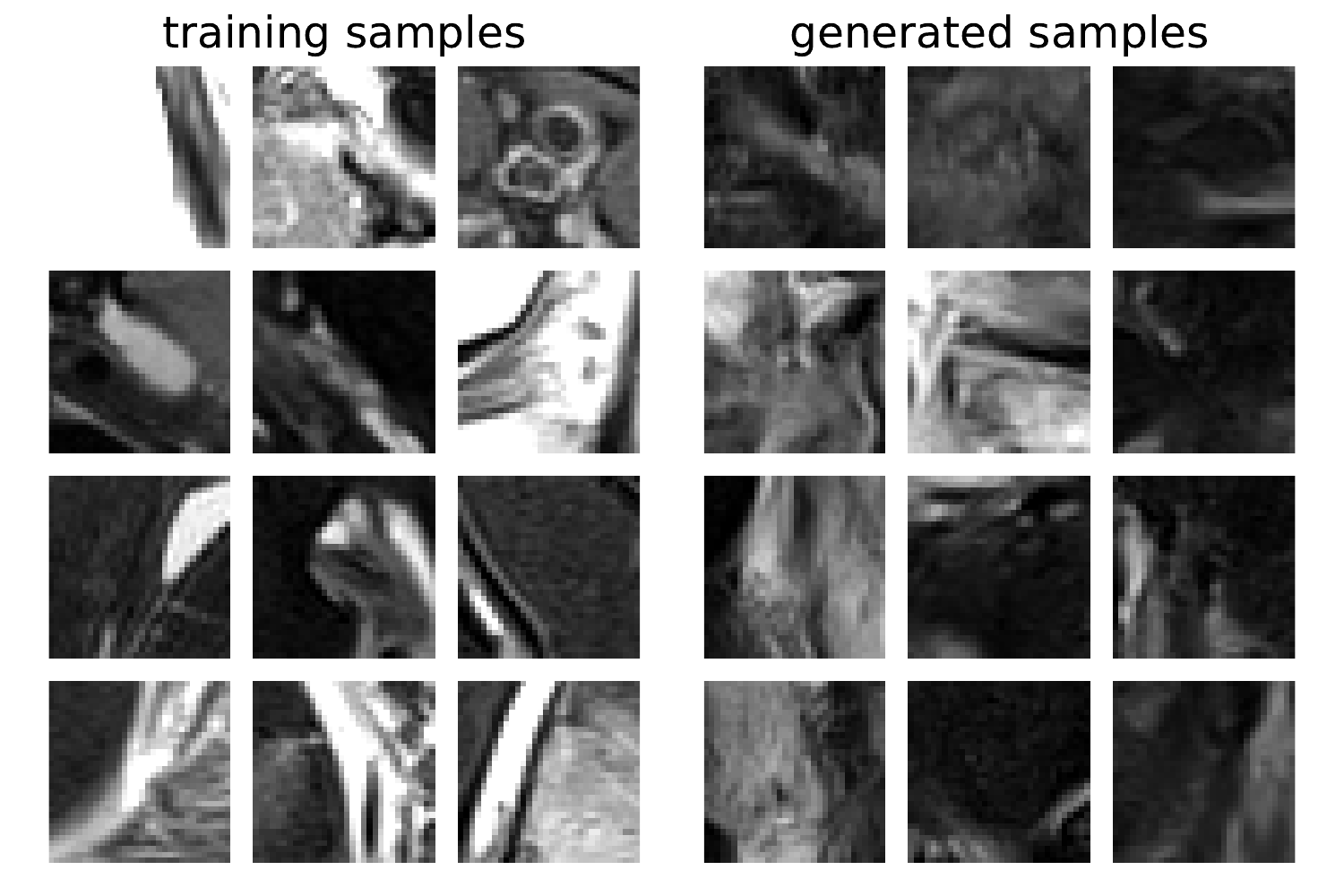}
    \caption{A set of patches sampled from the training data (left), and generated using the adopted normalizing flow model (right)}
    \label{fig:train_batch}
\end{figure}

As mentioned, we train our deep generative prior on patches sampled from high-quality MRI images, obtained from the NYU fast-MRI dataset (fully-sampled images of human knees) \cite{fastmri1, fastmri2}. During our initial experiments we however noticed that merely sampling patches at random from the full MRI volumes did not yield satisfactory results. The generative model was not sufficiently incentivised to model samples with strong inter-patch diversity and important features such as sharp tissue boundaries and fine structural details, since large parts of the MRI volumes did not contain such features. To promote structural fidelity and expressiveness of the model, we favour patches with sharp features during sampling, measured using the Tenengrad focus measure $\phi$ \cite{focus_measures}:
\begin{equation}
\phi = \sum_{(i, j) \in \Omega(x, y)}(G_x(i, j)^2 + G_y(i, j)^2),
\end{equation}
where $G_x(i, j)$, and $G_y(i, j)$ are the horizontal and vertical gradients computed by convolving the image with the horizontal and vertical Sobel filters respectively \cite{sobel}. Fig.~\ref{fig:train_batch} (left) shows a random batch of selected patches. The final train dataset consisted of $10^6$ such image patches. Note that for inference on simulations with many regions without texture (e.g. an artificial cyst phantom) which we complemented this with $50.000$ synthesized uniform (no texture) patches with intensities drawn randomly between $0$ and $0.5$. These additional patches are included to also enable assigning high likelihood to image regions that contain no texture. 



\begin{figure}[b!]
    \centering
    \includegraphics[trim={0cm 0cm 0cm 0},clip,width=1\linewidth]{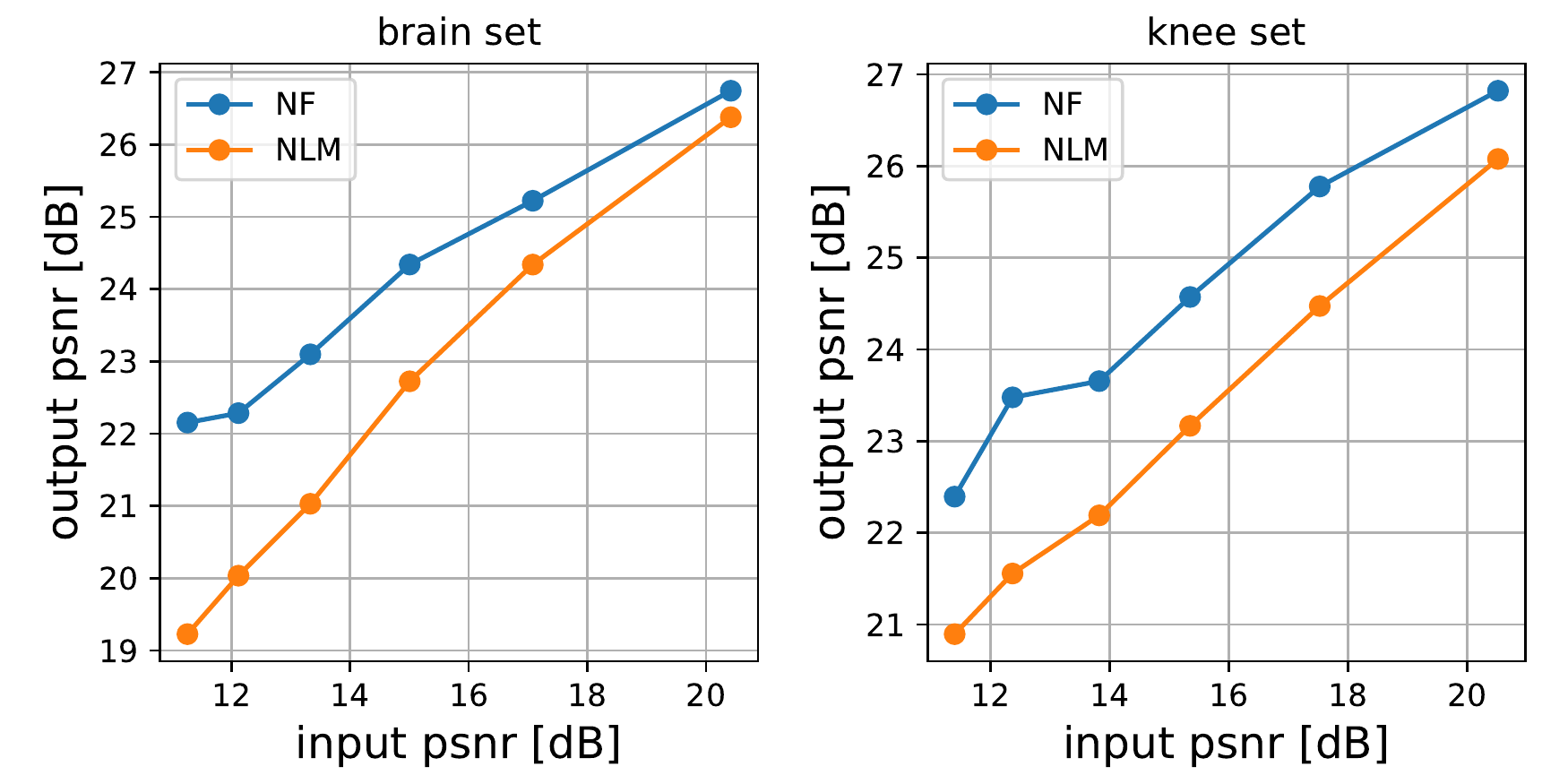}
    \includegraphics[trim={0cm 0cm 0cm 0},clip,width=1\linewidth]{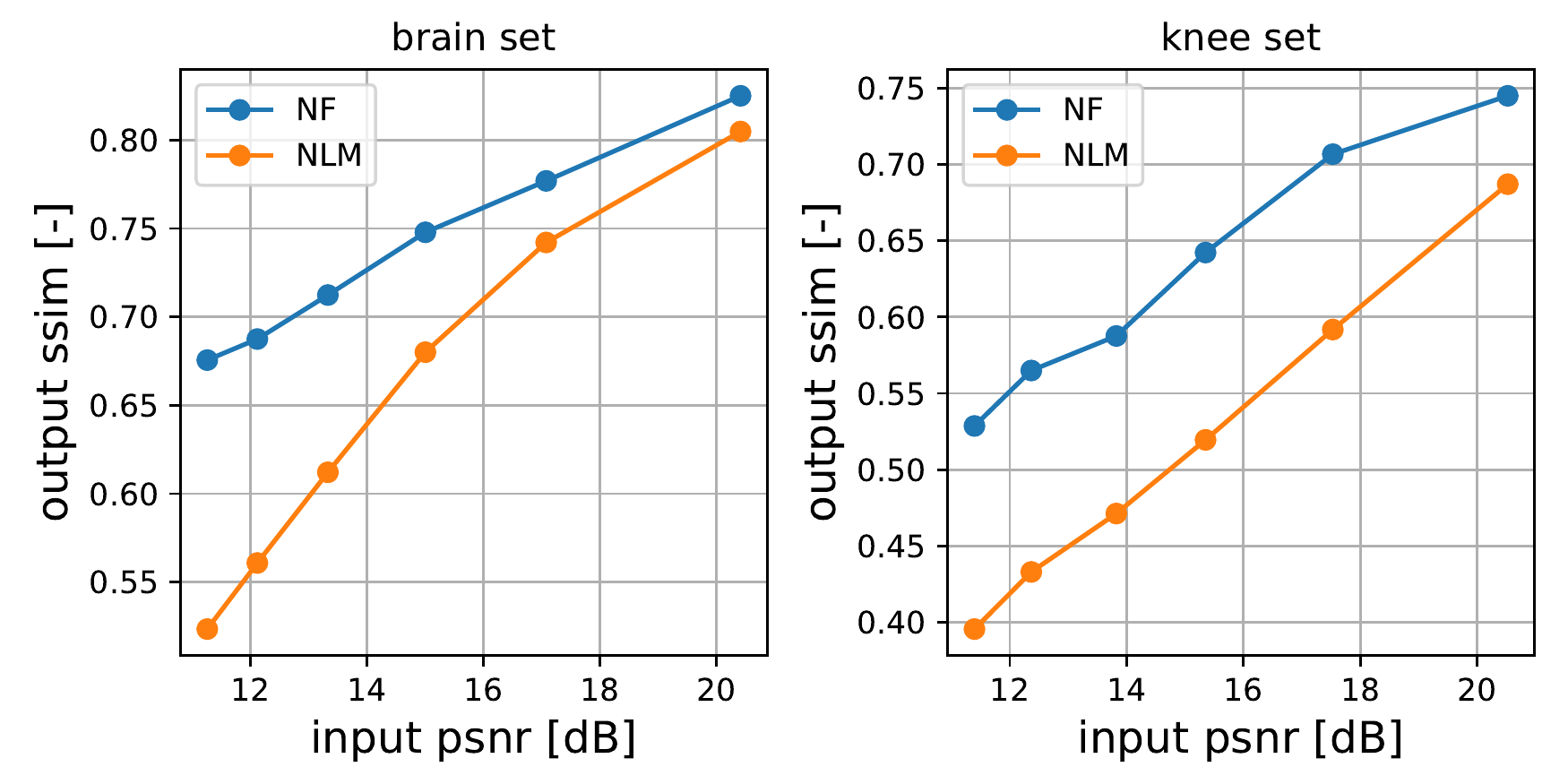}
    \caption{PSNR and SSIM across various Gaussian noise levels for images from both the knee and brain NYU fast-MRI datasets. Our method outperforms the standard non-local means filter (NLM), a competitive baseline in ultrasound image de-speckling that we optimized to maximize PSNR via a grid search.}.
    \label{fig:quantitativeresultGaussian}
\end{figure}

\begin{figure*}
    \centering
    \includegraphics[trim={0cm 8cm 0cm 0},clip,width=0.75\linewidth]{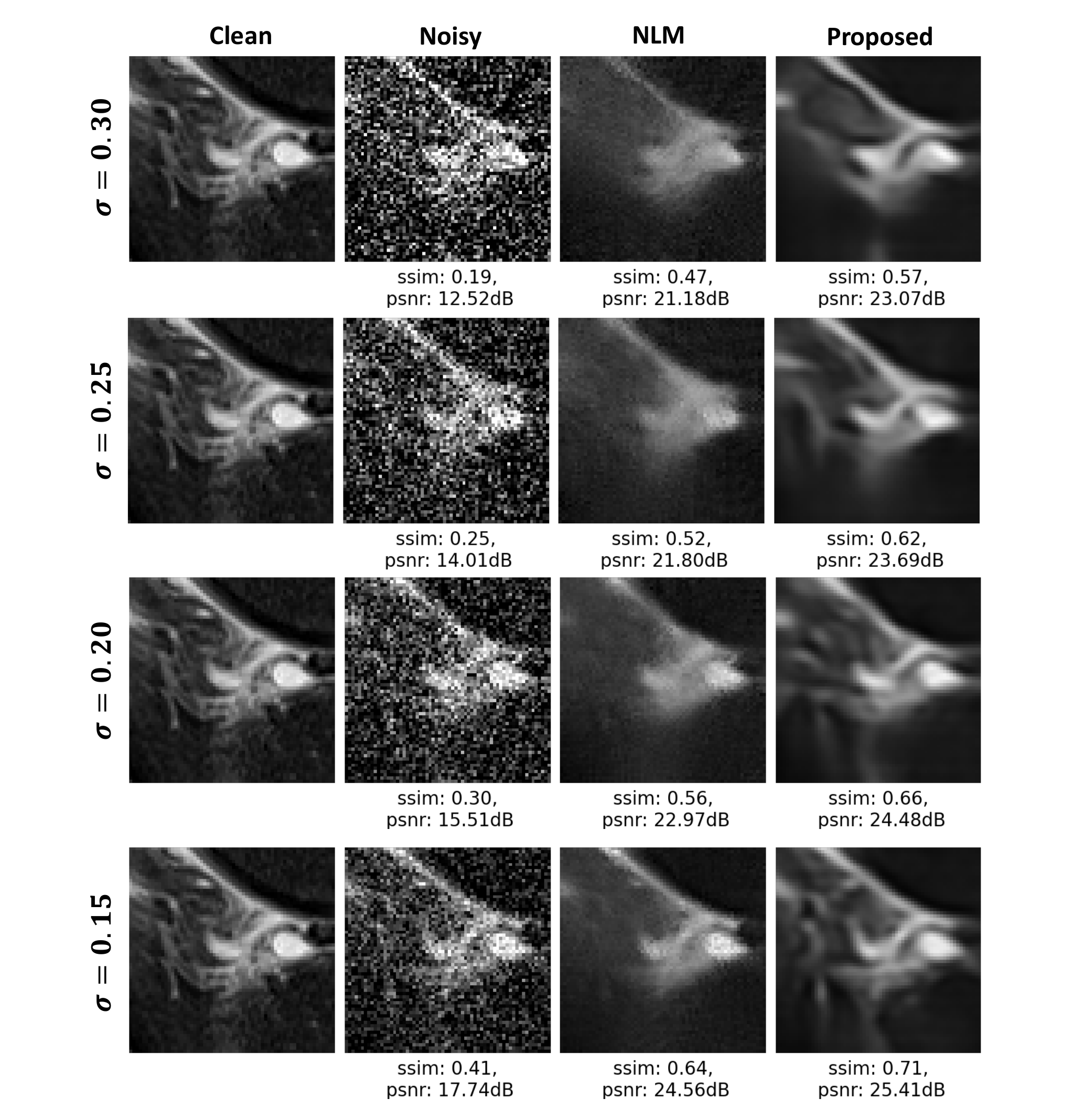}
    \caption{An illustrative example of denoising across various noise levels, comparing clean images, (optimized) \gls{NLM}, and the proposed method. The corresponding \gls{PSNR} and \gls{SSIM} with respect to the clean image are displayed below the images.}.
    \label{fig:resultGaussian}
\end{figure*}

\subsection{Normalizing flow architecture}
\label{section:methods_prior}
\noindent Our parameterization for the prior distribution is based on the Glow architecture \cite{glow}. 
Glow consists of a sequence of invertible steps, that each transform and mix the input dimensions. Each step comprises three consecutive layers: An actnorm layer, a 1x1-convolution layer, and an affine coupling layer.
The actnorm layer is performs an affine transformation that is initialized to result in a zero mean, and unit variance output. After initialization the parameters become free trainable parameters.

The 1x1-convolution layer can be seen as a generalisation of a permutation operation of the input channels. It mixes together the channels of the input by matrix-multiplying the vector of channel values of each pixel with a learned weight matrix $W$. $W$ is initialized as an orthogonal matrix.

Finally the data is passed through an affine coupling layer. The affine coupling layer splits up the input $\boldsymbol{a}$ into two equal parts, $\boldsymbol{a}_1$, and $\boldsymbol{a}_2$, along the channel dimension. It then performs an affine transformation on $\boldsymbol{a}_1$, where the applied scale and shift are a function of $\boldsymbol{a}_2$. $\boldsymbol{a}_2$ passes through unchanged. The two parts are then concatenated into output $\boldsymbol{b}$.
\begin{gather*}
    \boldsymbol{a}_1,\boldsymbol{a}_2 = \text{split}(\boldsymbol{a})\\
    (\log s, t) = \text{NN}(\boldsymbol{a}_2) \\
    s = S(\log s+2) \\
    \boldsymbol{b}_1 = s\cdot\boldsymbol{a}_1 + t \\
    \boldsymbol{b}_2 = \boldsymbol{a}_2 \\
    \boldsymbol{b} = \text{concat}(\boldsymbol{b}_2, \boldsymbol{b}_2),
\end{gather*}
where $S(\cdot)$ denotes the sigmoid function. 
The key advantage of this scheme is that for the coupling layer to be invertible, the function of $\boldsymbol{a}_2$, $NN(\boldsymbol{a}_2)$, does not have to be invertible. It can thus be an arbitrarily complex function or any neural network. Here we use the same convolutional neural network with three convolutional layers that was used in the Glow paper\cite{glow}.

The model was implemented in Pytorch and was trained with a batch size of $800$ for $10.000$ batches using the Adam optimizer. We use gradient checkpointing to reduce its memory footprint during training and inference \cite{gradient_checkpointing}. During training we moreover clip the gradients and the norm of all gradients to improve stability.\\

\section{Experiments}
\label{section:experiments}
We perform a series of experiments that incrementally close the gap between the assumptions made in the MAP estimator on noise statistics and real data. To that end we test performance on anatomical MRI samples with synthetic additive Gaussian noise, MRI-based synthetic US speckle simulations, and finally \textit{in-silico} and \textit{in-vivo} US data.

\begin{figure*}[t!]
    \centering
    \includegraphics[trim={0cm 22cm 0cm 0},clip,width=1\linewidth]{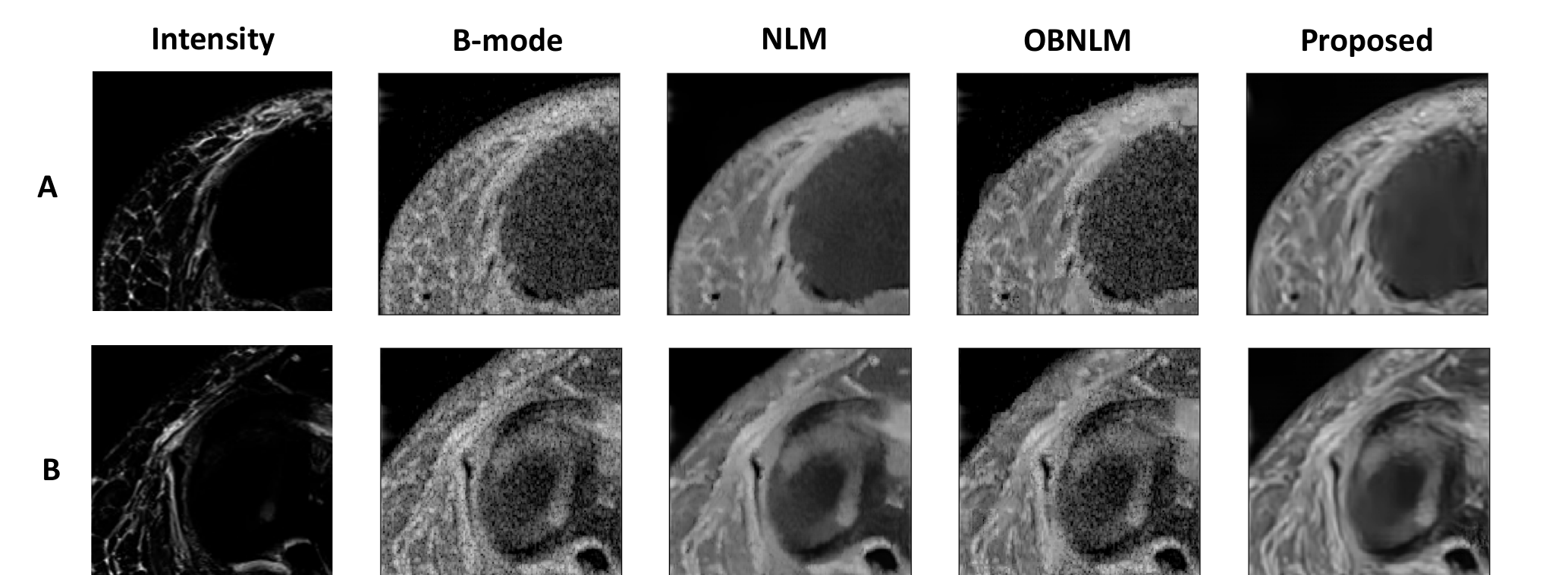}
    \caption{Speckle suppression on simulated ultrasound B-mode images from MRI-derived tissue backscatter intensity maps, comparing NLM, OBNLM (both qualitatively optimized through a grid search), and the proposed method.}.
    \label{fig:resultspeckle}
\end{figure*}

\subsection{Baselines}
\noindent Popular techniques for edge-preserving ultrasound speckle-suppression include anisotropic diffusion methods \cite{yu2002speckle} and stochastic iterative techniques for outlier removal \cite{tay2010ultrasound}. More recently, non-local means methods have demonstrated excellent performance in speckle reduction \cite{breivik2017real,ambrosanio2020wksr}. We thus compare our method with the standard Non-Local Means (NLM) \cite{nlm} and Optimized Bayesian Non-Local Means (OBNLM) \cite{obnlm} methods. Non-local means methods have been shown to outperform other ultrasound de-speckling algorithms \cite{coupe2008bayesian,ambrosanio2020wksr}, and are to date competitive \cite{Hyun2019,ambrosanio2020wksr}. For fair comparison we fine-tune the settings of these baseline algorithms to the data in the particular experiment via an exhaustive parameter grid search. We refer the reader to the original papers for more details.\\

Although recent speckle suppression based on end-to-end supervised training of deep neural networks has been proposed \cite{Hyun2019}, we here restrict ourselves to the unsupervised setting, i.e. the setting where no pairs of speckled and speckle-free images are available. 

\begin{figure*}
    \centering
    \includegraphics[trim={0cm 18.2cm 1cm 0},clip,width=\linewidth]{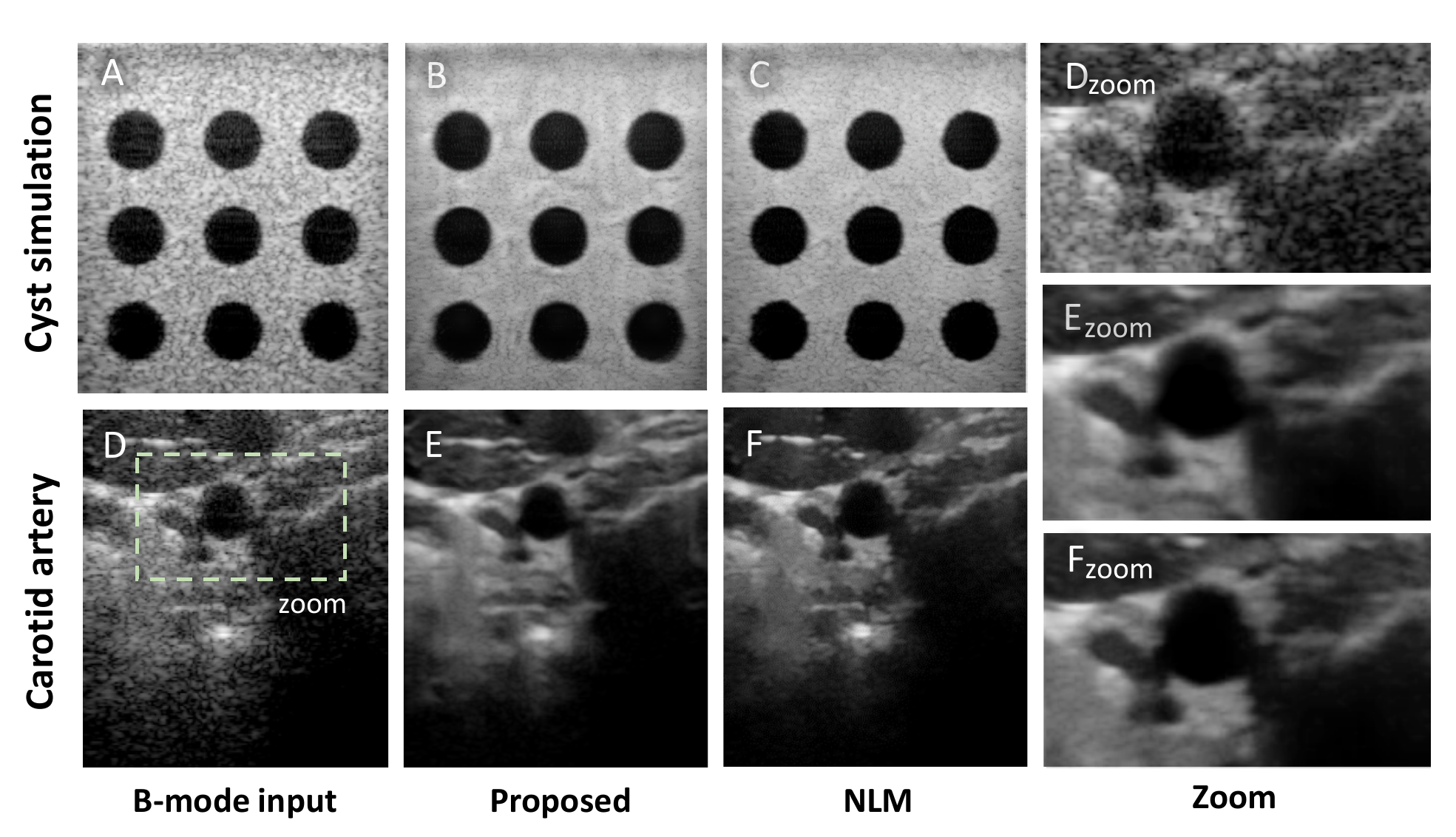}
    \caption{Comparison of the proposed method (B,E) against an optimized NLM baseline (C,F) on an artificial cyst simulation (top) and a human \textit{in-vivo} carotid artery (bottom). Note that we here mix back a small amount of speckle (20\%) for improved perceptual quality (as is common practice in clinical systems). Qualitatively, NLM produces patchy images \textit{in vivo}, giving a ``posterization'' effect, while the proposed method qualitatively preserves a more natural appearance an better reveals details. This is also apparent from the zoomed inset.}
    \label{fig:picmus_denoising}
\end{figure*}

\subsection{Denoising simulations}
\label{section:methods_experiments_simple}
\noindent We first test feasibility on a set of $16$ artificially noise-corrupted MRI images of $(64\times 64)$ pixels, extracted from a hold-out test subset of the NYU fastMRI database, containing knees and brains. Note that our NF prior is only trained on patches from images of knees, so the latter is useful for assessing performance on out-of-distribution anatomy. We apply additive Gaussian white noise having standard deviations  $\sigma_n\in\{0.1, 0.2, 0.3, 0.4\}$. For denoising we use a $\lambda=0.75\sigma^2$. Testing the algorithm for a range of noise intensities is important since speckle noise intensity in the image domain is known to be signal dependent (see Sec. \ref{section:background}).  
We compare ourselves to the \gls{NLM}-algorithm by evaluating the \gls{PSNR} and \gls{SSIM} statistics \cite{ssim}. We do not compare to \gls{OBNLM} here since it is not designed for additive Gaussian noise.

\subsection{Speckle suppression simulations}
\noindent To synthesize images with more realistic speckle noise from clean (MRI-based tissue contrast) images, we follow the method by Yu et al. in \cite{yu2002speckle}. Given an oversampled and normalized MRI tissue contrast image $\x$, we generate a synthetic image with speckle $\y$ through:
 \begin{equation}
    \y(r_x,r_y) = \left\lbrace \x(r_x,r_y)\cdot \n(r_x,r_y)\right\rbrace*h(r_x,r_y),
\end{equation}
with $\n(r_x,r_y)$ being Rayleigh-distributed noise and where 
\begin{equation}
    h(r_x,r_y) = \sin\left[\frac{2\pi f_0 r_x}{c}\right] \exp \left[-\frac{r_x^2}{2\sigma_{r_x}^2}\right]\exp \left[-\frac{r_y^2}{2\sigma_{r_y}^2}\right],
\end{equation}
with $f_0$ being the RF pulse center frequency, $c$ the speed of sound in the medium,  and $\sigma_{r_x}$ and $\sigma_{r_y}$ determining the axial and lateral resolution, respectively, of the virtual system. In our experiments we use $f_0 = 7$~MHz, $c=1540$~m/s, $\sigma_{r_x}=0.25$~mm and $\sigma_{r_y}=0.3$~mm. After this we perform envelope detection via the Hilbert transform, log compress the images, and subsequently downscale to the original image pixel dimensions. We again compare with the \gls{NLM} algorithm, and additionaly with the \gls{OBNLM} algorithm based on parts of knee MRI images that were withheld from the training set. 

\subsection{PICMUS and CUBDL datasets}
\label{section:methods_experiments_picmus}
\noindent We then evaluate performance on various US images from the \textit{\gls{PICMUS}} \cite{picmus} and \textit{\gls{CUBDL}} datasets \cite{CUBDL, rindal2017hypothesis, cubdl_data}. From the former we include images of a cyst phantom and a carotid artery; from the latter images of the heart. Before processing, the beamformed images were re-sampled and interpolated to obtain images having equal axial and lateral resolution and a pixel size that is about same as that of the MRI patches the generative model was trained on. For the cardiac images we performed denoising in the polar domain, i.e. before scan conversion. 
Finally the images were converted to decibels and normalized between zero and one. 
For denoising we used a fixed $\lambda=10$.

The NLM algorithm was applied with $h=0.8\sigma$, a patch size of $10$, and a patch distance of $10$. The OBNLM algorithm was applied with $h=\sigma$, a patch size of $3$, and a similarity window of size $10$.


\section{Results}
\label{section:results}
\subsection{Denoising simulations}
\label{section:results_simple}
\noindent We compare performance of the proposed method against our baseline at various noise levels in Fig.~\ref{fig:quantitativeresultGaussian}. The proposed method outperforms the NLM algorithm across all evaluated noise levels. Noteably, it also generalizes well to images from the brain dataset of the NYU fast-MRI challenge. This supports our hypothesis that by choosing a prior factorization that only models joint pixel dependencies across a relatively small local neighborhood of $1.4\times1.4$~cm, the model generalizes across anatomies. 

Fig.~\ref{fig:resultGaussian} qualitatively shows how performance is impacted by the overall noise level for the two methods in a typical example. Across the noise levels, the proposed method better captures the details and structure of the original image. Although we notice that our method produces slightly more blurred features compared to the clean input, it visually outperforms the NLM algorithm. \\



\subsection{Speckle suppression simulations}
\label{section:results_specklesimulations}
\noindent Fig.~\ref{fig:resultspeckle} shows the results on two speckle simulations generated using intensity maps derived from MRI images of (segments of) human knees. Comparing the resulting denoised images by NLM, OBNLM and the proposed methods with the intensity maps (and its features), we observe that both NLM and the proposed method provide strong denoising. OBNLM preserves more details than NLM, but also retains more noise. The proposed method preserves details well, without sacrificing denoising capabilities. 

\subsection{PICMUS and CUBDL datasets}
\label{section:results_picmus}
\noindent The results from experiments on the \gls{PICMUS} dataset are shown in Fig.~\ref{fig:picmus_denoising}. The first observation is that the proposed method translates well to the real \textit{in-vivo} ultrasound data, showing the feasibility of using MRI-based image priors for ultrasound denoising and speckle suppression. While the MRI database did not include perfect circle-shaped low-intensity regions, the proposed method does not distort the shapes and remains true to the geometry, showing the strong generalization of our prior.  Moreover, comparing to the NLM method, we observe that it produces more natural and less patchy denoising, as is common with NLM methods. Although this effect is not very noticeable on the cyst simulations since these artificial constellations indeed only consist of two levels of scattering intensity, it is clear on the more intricate \textit{in-vivo} data of a human carotid artery. From the zoomed areas we can also appreciate better preservation of detail with the proposed approach. 

\begin{figure*}
    \centering
    \includegraphics[trim={0cm 23.5cm 0cm 0},clip, width=\linewidth]{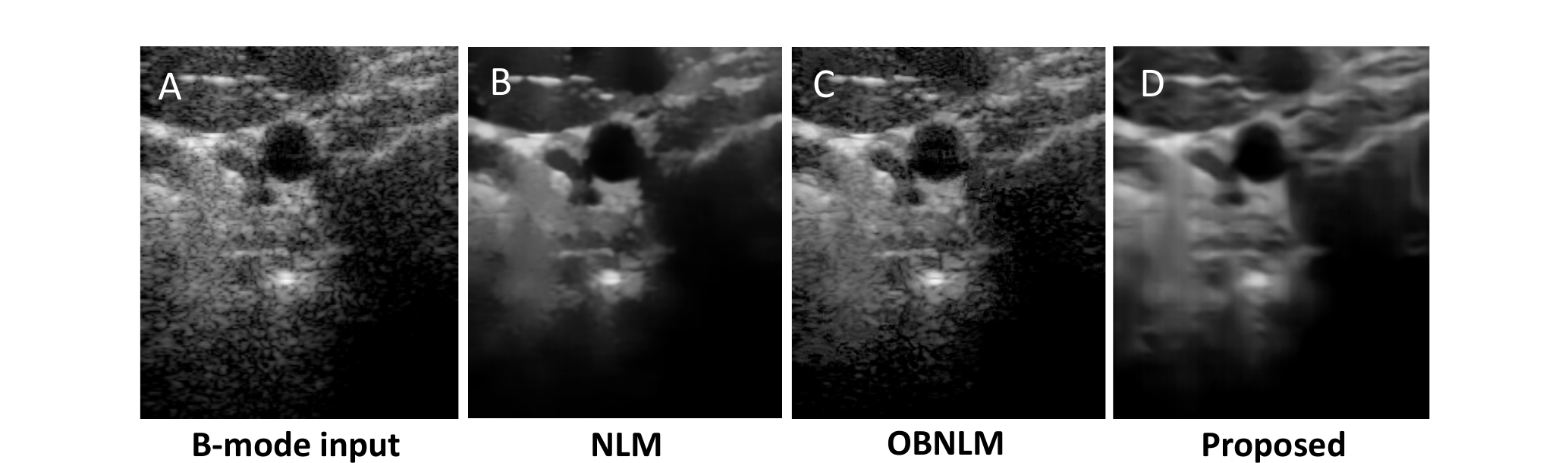}
    \caption{Comparison among NLM (B), OBNLM (C), and the proposed method (D) \textit{in-vivo}. Qualitatively, OBNLM is less patchy than NLM, but also more noisy. The proposed method retains a more natural appearance, yet with strong noise suppression.}
    \label{fig:extended_comparison}
\end{figure*}

Fig.~\ref{fig:extended_comparison} offers a side by side comparison including OBNLM. OBNLM has been designed to cope with the intensity-dependent noise statistics, and produces less patchy results. Yet, OBNLM qualitatively also remains much more noisy.

\begin{figure*}[t!]
    \centering
    \includegraphics[trim={0cm 17cm 0cm 0},clip, width=0.8\linewidth]{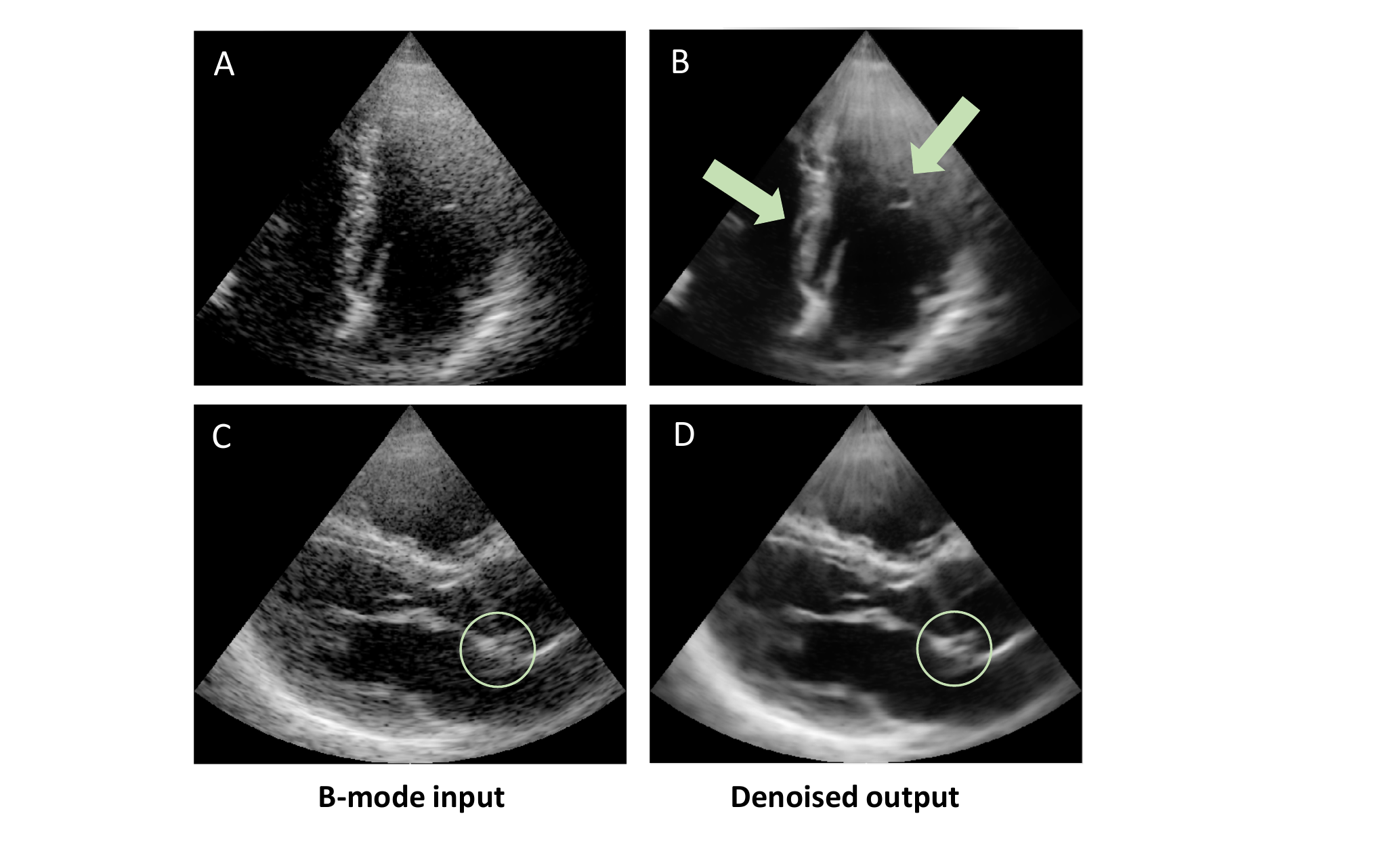}
    \caption{Additional denoising results on cardiac images from the CUBDL dataset \cite{cubdl_data}, demonstrating how the method generalizes across various anatomies.}
    \label{fig:comparison_osl}
\end{figure*}

We show some additional denoising results on cardiac images from the CUBDL database in Fig.~\ref{fig:comparison_osl}. This indicates how the method generalizes across domains, and effectively leverages the image priors derived from MRI for various applications.   \\

\section{Discussion}
\label{section:discussion}

\noindent In this paper, we demonstrated that patch-based iterative denoising with a deep generative normalizing-flow image prior can produce sharp and detailed images from ultrasound acquisitions distorted by severe noise and speckle. Our experiments show that the proposed method numerically outperforms a fine-tuned \gls{NLM} algorithm in terms of both \gls{SSIM}, and \gls{PSNR} across a range of noise powers. The method qualitatively outperforms both \gls{NLM} and \gls{OBNLM} on ultrasound speckle simulations and real \textit{in-vivo} data from the PICMUS challenge. Notably, it produces less patchy and perceptually more natural outputs compared to the other algorithms, which yield an almost segmented output, reminiscent of ``posterization''.

Although the results are promising, the required processing time used for iterative MAP inference is currently too high for real-time application. Although not the focus of this work, one avenue to explore is the use of deep unfolding \cite{monga2021algorithm}, in which the iterations of the algorithm are unfolded as a feed-forward deep network \cite{wei2021deep}. In other applications this has yielded major speedups (as high as a factor 100) \cite{solomon2019deep,van2019deep}, potentially enabling real-time inference. Another means of acceleration would be to exploit the temporal structure/persistence of ultrasound sequences, initializing the MAP solver for the next frame with the with the solution for the previous frame. This way the algorithm ``tracks'' the statistics. We leave this for future work. 

To further boost performance, one could explore improvements in the adopted data likelihood model. While we here assume white Gaussian error statistics, this is in practice naive, since speckle noise is (1) spatially correlated, and (2) signal-dependent. Formulating a more accurate model could lead in the future lead to additional performance gains. Note that such models can also be formulated to jointly perform denoising/despeckling and deconvolution, by including a (convolutional) point-spread function model in Eqn.~\ref{eqn:MAPfinal}.

More broadly, we foresee application of the proposed MRI-derived image priors to ultrasound compressed sensing settings such as sub-Nyquist acquisitions \cite{chernyakova2014fourier} and sparse arrays \cite{huijben2020learning}, e.g. overcoming the challenges of analytically formulating an adequate sparsifying basis with learned priors. 

Finally, a more in-depth comparison with a large set of real ultrasound images is required, including clinical assessment, which we leave to future work. 

\section{Conclusion}
\label{section:conclusion}
We present a new method for ultrasound image denoising and speckle suppression that uses MRI-based structural priors learned with deep generative models to perform maximum-a-posteriori estimation of clean images. To learn intricate image priors from MRI data that generalize to unseen anatomies, we devise an invertible normalizing flow architecture that models a factorized density function based on local image patches. Notably, the proposed deep learning method thus requires no supervision for its inference (no pairs of clean and noisy data), and outperforms strong baselines based on non-local-means quantitatively and qualitatively.

\bibliography{library}

\begin{thebibliography}{10}

\bibitem{lee_filter}
J.-S. Lee, ``Digital image enhancement and noise filtering by use of local
  statistics,'' 1977.

\bibitem{frost_filter}
V.~S. Frost, J.~A. Stiles, K.~S. Shanmugan, and J.~C. Holtzman, ``A model for
  radar images and its application to adaptive digital filtering of
  multiplicative noise,'' {\em IEEE Transactions on Pattern Analysis and
  Machine Intelligence}, vol.~PAMI-4, pp.~157--166, 1982.

\bibitem{kuan_filter}
D.~T. Kuan, A.~A. Sawchuk, T.~C. Strand, and P.~Chavel, ``Adaptive noise
  smoothing filter for images with signal-dependent noise.,'' {\em IEEE
  Transactions on Pattern Analysis and Machine Intelligence}, vol.~PAMI-7,
  pp.~165--177, 1985.

\bibitem{srad}
Y.~Yu and S.~T. Acton, ``Speckle reducing anisotropic diffusion,'' {\em IEEE
  Transactions on Image Processing}, vol.~11, pp.~1260--1270, 11 2002.

\bibitem{nlm}
A.~Buades, B.~Coll, and J.~M. Morel, ``A non-local algorithm for image
  denoising,'' vol.~II, pp.~60--65, IEEE Computer Society, 2005.

\bibitem{obnlm}
Y.~Zhan, M.~Ding, L.~Wu, and X.~Zhang, ``Nonlocal means method using weight
  refining for despeckling of ultrasound images,'' {\em Signal Processing},
  vol.~103, pp.~201--213, 2014.

\bibitem{coupe2008bayesian}
P.~Coup{\'e}, P.~Hellier, C.~Kervrann, and C.~Barillot, ``Bayesian non local
  means-based speckle filtering,'' in {\em 2008 5th IEEE International
  Symposium on Biomedical Imaging: From Nano to Macro}, pp.~1291--1294, IEEE,
  2008.

\bibitem{structural_convolutional}
D.~Li, W.~Yu, K.~Wang, D.~Jiang, and Q.~Jin, ``Speckle noise removal based on
  structural convolutional neural networks with feature fusion for medical
  image,'' {\em Signal Processing: Image Communication}, vol.~99, p.~116500, 11
  2021.

\bibitem{5_NN}
O.~Karaoğlu, H.~Şakir Bilge, and İhsan Uluer, ``Removal of speckle noises
  from ultrasound images using five different deep learning networks,'' {\em
  Engineering Science and Technology, an International Journal}, 2021.

\bibitem{NN_beamforming}
D.~Hyun, L.~L. Brickson, K.~T. Looby, and J.~J. Dahl, ``Beamforming and speckle
  reduction using neural networks,'' {\em IEEE Transactions on Ultrasonics,
  Ferroelectrics, and Frequency Control}, vol.~66, pp.~898--910, 5 2019.

\bibitem{vedula2017towards}
S.~Vedula, O.~Senouf, A.~M. Bronstein, O.~V. Michailovich, and M.~Zibulevsky,
  ``Towards ct-quality ultrasound imaging using deep learning,'' {\em arXiv
  preprint arXiv:1710.06304}, 2017.

\bibitem{feng2020ultrasound}
X.~Feng, Q.~Huang, and X.~Li, ``Ultrasound image de-speckling by a hybrid deep
  network with transferred filtering and structural prior,'' {\em
  Neurocomputing}, vol.~414, pp.~346--355, 2020.

\bibitem{luijten2020adaptive}
B.~Luijten, R.~Cohen, F.~J. de~Bruijn, H.~A. Schmeitz, M.~Mischi, Y.~C. Eldar,
  and R.~J. van Sloun, ``Adaptive ultrasound beamforming using deep learning,''
  {\em IEEE Transactions on Medical Imaging}, vol.~39, no.~12, pp.~3967--3978,
  2020.

\bibitem{dietrichson2018ultrasound}
F.~Dietrichson, E.~Smistad, A.~Ostvik, and L.~Lovstakken, ``Ultrasound speckle
  reduction using generative adversial networks,'' in {\em 2018 IEEE
  International Ultrasonics Symposium (IUS)}, pp.~1--4, IEEE, 2018.

\bibitem{tierney2020domain}
J.~Tierney, A.~Luchies, C.~Khan, B.~Byram, and M.~Berger, ``Domain adaptation
  for ultrasound beamforming,'' in {\em International Conference on Medical
  Image Computing and Computer-Assisted Intervention}, pp.~410--420, Springer,
  2020.

\bibitem{realnvp}
L.~Dinh, J.~Sohl-Dickstein, and S.~Bengio, ``Density estimation using real
  nvp,'' 5 2016.

\bibitem{nice}
L.~Dinh, D.~Krueger, and Y.~Bengio, ``Nice: Non-linear independent components
  estimation,'' 10 2014.

\bibitem{glow}
D.~P. Kingma and P.~Dhariwal, ``Glow: Generative flow with invertible 1x1
  convolutions,'' 7 2018.

\bibitem{flow++}
J.~Ho, X.~Chen, A.~Srinivas, Y.~Duan, and P.~Abbeel, ``Flow++: Improving
  flow-based generative models with variational dequantization and architecture
  design,'' 2 2019.

\bibitem{overview1}
I.~Kobyzev, S.~J.~D. Prince, and M.~A. Brubaker, ``Normalizing flows: An
  introduction and review of current methods,'' 8 2019.

\bibitem{asim2020invertible}
M.~Asim, M.~Daniels, O.~Leong, A.~Ahmed, and P.~Hand, ``Invertible generative
  models for inverse problems: mitigating representation error and dataset
  bias,'' in {\em International Conference on Machine Learning}, pp.~399--409,
  PMLR, 2020.

\bibitem{wei2021deep}
X.~Wei, H.~van Gorp, L.~G. Carabarin, D.~Freedman, Y.~Eldar, and R.~van Sloun,
  ``Deep unfolding with normalizing flow priors for inverse problems,'' {\em
  arXiv preprint arXiv:2107.02848}, 2021.

\bibitem{maf}
G.~Papamakarios, T.~Pavlakou, and I.~Murray, ``Masked autoregressive flow for
  density estimation,'' 5 2017.

\bibitem{fastmri1}
J.~Zbontar, F.~Knoll, A.~Sriram, T.~Murrell, Z.~Huang, M.~J. Muckley,
  A.~Defazio, R.~Stern, P.~Johnson, M.~Bruno, M.~Parente, K.~J. Geras,
  J.~Katsnelson, H.~Chandarana, Z.~Zhang, M.~Drozdzal, A.~Romero, M.~Rabbat,
  P.~Vincent, N.~Yakubova, J.~Pinkerton, D.~Wang, E.~Owens, C.~L. Zitnick,
  M.~P. Recht, D.~K. Sodickson, and Y.~W. Lui, ``fastmri: An open dataset and
  benchmarks for accelerated mri,'' 11 2018.

\bibitem{fastmri2}
F.~Knoll, J.~Zbontar, A.~Sriram, M.~J. Muckley, M.~Bruno, A.~Defazio,
  M.~Parente, K.~J. Geras, J.~Katsnelson, H.~Chandarana, Z.~Zhang,
  M.~Drozdzalv, A.~Romero, M.~Rabbat, P.~Vincent, J.~Pinkerton, D.~Wang,
  N.~Yakubova, E.~Owens, C.~L. Zitnick, M.~P. Recht, D.~K. Sodickson, and Y.~W.
  Lui, ``fastmri: A publicly available raw k-space and dicom dataset of knee
  images for accelerated mr image reconstruction using machine learning,'' {\em
  Radiology: Artificial Intelligence}, vol.~2, p.~e190007, 1 2020.

\bibitem{yu2002speckle}
Y.~Yu and S.~T. Acton, ``Speckle reducing anisotropic diffusion,'' {\em IEEE
  Transactions on image processing}, vol.~11, no.~11, pp.~1260--1270, 2002.

\bibitem{gradient_checkpointing}
T.~Chen, B.~Xu, C.~Zhang, and C.~Guestrin, ``Training deep nets with sublinear
  memory cost,'' 4 2016.

\bibitem{focus_measures}
S.~Pertuz, D.~Puig, and M.~A. Garcia, ``Analysis of focus measure operators for
  shape-from-focus,'' {\em Pattern Recognition}, vol.~46, pp.~1415--1432, 5
  2013.

\bibitem{sobel}
I.~Sobel, ``An isotropic 3x3 image gradient operator,'' {\em Presentation at
  Stanford A.I. Project 1968}, 11 2014.

\bibitem{tay2010ultrasound}
P.~C. Tay, C.~D. Garson, S.~T. Acton, and J.~A. Hossack, ``Ultrasound
  despeckling for contrast enhancement,'' {\em IEEE Transactions on Image
  Processing}, vol.~19, no.~7, pp.~1847--1860, 2010.

\bibitem{breivik2017real}
L.~H. Breivik, S.~R. Snare, E.~N. Steen, and A.~H.~S. Solberg, ``Real-time
  nonlocal means-based despeckling,'' {\em IEEE transactions on ultrasonics,
  ferroelectrics, and frequency control}, vol.~64, no.~6, pp.~959--977, 2017.

\bibitem{ambrosanio2020wksr}
M.~Ambrosanio, B.~Kanoun, and F.~Baselice, ``wksr-nlm: An ultrasound
  despeckling filter based on patch ratio and statistical similarity,'' {\em
  IEEE Access}, vol.~8, pp.~150773--150783, 2020.

\bibitem{Hyun2019}
D.~Hyun, L.~L. Brickson, K.~T. Looby, and J.~J. Dahl, ``Beamforming and speckle
  reduction using neural networks,'' {\em IEEE Transactions on Ultrasonics,
  Ferroelectrics, and Frequency Control}, vol.~66, pp.~898--910, 5 2019.

\bibitem{ssim}
Z.~Wang, A.~C. Bovik, H.~R. Sheikh, and E.~P. Simoncelli, ``Image quality
  assessment: From error visibility to structural similarity,'' {\em IEEE
  Transactions on Image Processing}, vol.~13, pp.~600--612, 4 2004.

\bibitem{picmus}
H.~Liebgott, A.~Rodriguez-Molares, F.~Cervenansky, J.~A. Jensen, and
  O.~Bernard, ``Plane-wave imaging challenge in medical ultrasound,''
  vol.~2016-November, IEEE Computer Society, 11 2016.

\bibitem{CUBDL}
D.~Hyun, A.~Wiacek, S.~Goudarzi, S.~Rothlubbers, A.~Asif, K.~Eickel, Y.~C.
  Eldar, J.~Huang, M.~Mischi, H.~Rivaz, D.~Sinden, R.~J.~V. Sloun, H.~Strohm,
  and M.~A.~L. Bell, ``Deep learning for ultrasound image formation: Cubdl
  evaluation framework and open datasets,'' {\em IEEE Transactions on
  Ultrasonics, Ferroelectrics, and Frequency Control}, 12 2021.

\bibitem{rindal2017hypothesis}
O.~M.~H. Rindal, S.~Aakhus, S.~Holm, and A.~Austeng, ``Hypothesis of improved
  visualization of microstructures in the interventricular septum with
  ultrasound and adaptive beamforming,'' {\em Ultrasound in Medicine \&
  Biology}, vol.~43, no.~10, pp.~2494--2499, 2017.

\bibitem{cubdl_data}
M.~A.~L. Bell, J.~Huang, A.~Wiacek, P.~Gong, S.~Chen, A.~Ramalli, P.~Tortoli,
  B.~Luijten, M.~Mischi, O.~M.~H. Rindal, V.~Perrot, H.~Liebgott, X.~Zhang,
  J.~Luo, E.~Oluyemi, and E.~Ambinder, ``{Challenge on Ultrasound Beamforming
  with Deep Learning (CUBDL) Datasets}.''

\bibitem{monga2021algorithm}
V.~Monga, Y.~Li, and Y.~C. Eldar, ``Algorithm unrolling: Interpretable,
  efficient deep learning for signal and image processing,'' {\em IEEE Signal
  Processing Magazine}, vol.~38, no.~2, pp.~18--44, 2021.

\bibitem{solomon2019deep}
O.~Solomon, R.~Cohen, Y.~Zhang, Y.~Yang, Q.~He, J.~Luo, R.~J. van Sloun, and
  Y.~C. Eldar, ``Deep unfolded robust pca with application to clutter
  suppression in ultrasound,'' {\em IEEE transactions on medical imaging},
  vol.~39, no.~4, pp.~1051--1063, 2019.

\bibitem{van2019deep}
R.~J. Van~Sloun, R.~Cohen, and Y.~C. Eldar, ``Deep learning in ultrasound
  imaging,'' {\em Proceedings of the IEEE}, vol.~108, no.~1, pp.~11--29, 2019.

\bibitem{chernyakova2014fourier}
T.~Chernyakova and Y.~C. Eldar, ``Fourier-domain beamforming: the path to
  compressed ultrasound imaging,'' {\em IEEE transactions on ultrasonics,
  ferroelectrics, and frequency control}, vol.~61, no.~8, pp.~1252--1267, 2014.

\bibitem{huijben2020learning}
I.~A. Huijben, B.~S. Veeling, K.~Janse, M.~Mischi, and R.~J. van Sloun,
  ``Learning sub-sampling and signal recovery with applications in ultrasound
  imaging,'' {\em IEEE Transactions on Medical Imaging}, vol.~39, no.~12,
  pp.~3955--3966, 2020.

\end{thebibliography}
\bibliographystyle{ieeetr}

\end{document}